\documentstyle[twoside,fleqn,espcrc2]{article}

\newcommand{\be}{\begin{equation}}
\newcommand{\ee}{\end{equation}}
\newcommand{\bea}{\begin{eqnarray}}
\newcommand{\eea}{\end{eqnarray}}

\newcommand{\AmS}{{\protect\the\textfont2
  A\kern-.1667em\lower.5ex\hbox{M}\kern-.125emS}}

\newcommand{\twooneplaq}{\setlength{\unitlength}{.5cm}
   \raisebox{-.2cm}{
   \begin{picture}(2.2,1.2)(-1.1,-.6)
   \put(-1,-.5){\line(1,0){2}}
   \put(-1,.5){\line(1,0){2}}
   \put(-1,-.5){\line(0,1){1}}
   \put(1,-.5){\line(0,1){1}}
   \put(-1,-0.5){\circle*{.2}}
   \put(-1.5,-0.9){$x$}
   \put(-0.05,-1.2){$\mu$}
   \put(-1.5,-0.05){$\nu$}
   \end{picture}}}
\newcommand{\ltwooneplaq}{\setlength{\unitlength}{.5cm}
   \raisebox{-.2cm}{
   \begin{picture}(2.2,1.2)(-1.1,-.6)
   \put(-.5,-1){\line(1,0){1}}
   \put(-.5,1){\line(1,0){1}}
   \put(-.5,-1){\line(0,1){2}}
   \put(.5,-1){\line(0,1){2}}
   \put(-0.5,-1){\circle*{.2}}
   \put(-0.9,-1.5){$x$}
   \put(-1.2,-0.05){$\nu$}
   \put(-0.05,-1.5){$\mu$}
   \end{picture}}}
\def\tint{\tau_{int}}

\title{Noisy Monte Carlo Algorithm
\thanks{A more detailed description of the NMC algorithm can be found in Ref.\cite{BF}}}

\author{T. Bakeyev\address{Joint Inst. for Nuclear Research, 
                             141980 Dubna, Russia}
                  \thanks{Talk presented by T.B. at Lattice 2000, Bangalore, India}          
    and Ph. de Forcrand
                  \address{Inst. f\"ur Theoretische Physik, 
                  ETH-H\"onggerberg, CH-8093 Z\"urich, Switzerland}
                  \address{CERN, Theory Division, CH-1211 Geneva 23, Switzerland}
       }

\begin{document}

\begin{abstract}
We present an exact Monte Carlo algorithm designed to sample theories where
the energy is a sum of many couplings of decreasing strength.
The algorithm avoids
the computation of almost all non-leading terms. Its use is illustrated
by simulating SU(2) lattice gauge theory with a 5-loop improved action. 
A new approach for dynamical fermion simulations is proposed.
\vspace{1pc}
\end{abstract}

\maketitle

When sampling the partition function 
$Z = \int \prod dU ~~ e^{-H(\{U\})}$,
the most common algorithm is that of Metropolis:
at each step, starting from the configuration $\{U\}$, a candidate
configuration $\{U'\}$ is proposed, and it is accepted with probability
$P_{acc} = \rm{min}(1, e^{- (H(\{U'\}) - H(\{U\}))} )$.
This acceptance test is realized by comparing $P_{acc}$
to a random number uniformly distributed in $[0,1]$. This seems like an excess of
information: why compute $H(\{U'\})$ {\em exactly}, then compare it with a random
number? It should be sufficient to {\em estimate} it. This logical 
proposition was studied long ago \cite{KK,BK}.
The main difficulty which prevented the construction of an efficient algorithm 
at that stage was the existence of probability bound violations for  
the noisy estimator of $P_{acc}$, which caused intolerable systematic errors. 
This difficulty was overcome in \cite{KFLiu}. Ref.\cite{KFLiu}, however, introduces
an infinite number of auxiliary variables, and tests of the 
method are performed on a toy model with 5 degrees of freedom only.
Here, we simplify the method of \cite{KFLiu},
by introducing only 1 auxiliary variable per term in $H$. Moreover, we separate
$H$ into a leading part to be calculated exactly, and a sum of correction
terms, which we treat stochastically. 

Consider a generic Hamiltonian of the type $H = \sum_{k=0}^m c_k W_k$,
where as $k$ increases, $|c_k|$ decreases. 
It is often the case that one would like to study a Hamiltonian of such type
resulting from an expansion, be it perturbative \cite{Symanzik}, non-perturbative
\cite{Lepage}, or based on the fixed point of a renormalization group 
transformation \cite{Hasenfratz}.
As $k$ increases, the number of geometrically equivalent terms
grouped into $W_k$ increases exponentially.
This combinatoric explosion normally makes
the simulation of extended Hamiltonians prohibitively expensive. 
However, in most cases the couplings $c_k$ decrease exponentially
with $k$, so that the overall Hamiltonian is dominated by $W_0$.
By making use of stochastic methods to estimate the
correction terms $W_k, k \ge 1$, we aim at postponing the combinatoric explosion
of the simulation costs incurred when including higher terms $W_k$. This opens
the possibility of studying numerically much more complicated Hamiltonians
including higher-order correction terms.

By shifting the $W_k$'s by irrelevant constants, one can arrange that
the terms $c_kW_k$
be nonpositive starting from $k=1$:
$c_k W_k(U)\le 0 ~~ \forall U$.
The contribution of the terms
$W_k(U), k\ge 1$ is estimated stochastically by introducing auxiliary fields
$\sigma_k$, which can take two values: 0 and 1. We represent the probability 
$e^{-H}$ in the form $P_0[U] *  P_1[U,\sigma]$, where
\bea \lefteqn{P_0[U]=e^{-c_0W_0(U)}} \nonumber \\   
\nopagebreak
\lefteqn{P_1[U,\sigma]\! =\! \prod_{k=1}^{m}
\sum_{\sigma_k=0,1}\!\![\delta_{\sigma_k,0}+\delta_{\sigma_k,1}(e^{-c_kW_k}-1)]}
\label{sigpr} \eea
$P_0[U] *  P_1[U,\sigma]$ can be interpreted as the joint probability distribution for the
original fields of the model and the new $\sigma$ fields.

One can easily see why the introduction of auxiliary $\sigma$ fields can be useful.
Starting from the configuration $\{ U_1,\sigma\}$, a candidate configuration
$\{ U_2\}$ distributed with the weight $P_0[U_2]$ is proposed, and accepted with
probability 
\bea 
\lefteqn{P_{acc} = \rm{min}\Bigl( 1, \frac{P_1[U_2,\sigma]}{P_1[U_1,\sigma]} \Bigr) =}
\nonumber \\ 
\lefteqn{= \rm{min}\Bigl( 1, \prod_{k\; :\;\sigma_k=1}\frac{e^{-c_kW_k(U_2)}-1}{e^{-c_kW_k(U_1)}-1}
\Bigr) }
\label{pacc} \eea
Since the terms $c_k W_k(U)$ contribute in $P_{acc}$ only if $\sigma_k=1$, the amount of 
computational work
is greatly reduced if the configurations with $\sigma_k=0$ are dominating. That is certainly 
the case when the coefficients $|c_k|$ are small: 
the probabilities for $\sigma_k$ to be unity are negligible then:
$p_{\sigma_k=1}=1-e^{c_kW_k(U)}
\approx |c_k W_k(U)|\approx 0\quad  {\rm if}\quad c_k\approx 0$.

We illustrate these ideas on a 5-loop perturbatively improved SU(2) gauge model in $4d$:
\bea
\lefteqn{S = \sum_{i=1}^5 c_i {1 \over {m_i^2 n_i^2}} S_{m_i,n_i} }
\nonumber \\
\lefteqn{S_{m_i,n_i}=} \nonumber \\ 
\lefteqn{\sum_{x,\mu,\nu} 
\left( -2{\rm sgn\; (c_i)} -\frac{{\rm Tr}}{2} \
 ( \   \twooneplaq + \hspace{-.1cm}\ltwooneplaq\ \hspace{-.1cm} ) \right)}
\label{5Li}\eea
where $(m_i,n_i) = (1,1),(2,2),(1,2),(1,3),(3,3)$ 
for $i=1,\ldots , 5$ denote the planar, fundamental loops of size $m\times n$.
The Gibbs factor is $\ {\rm exp}(- \frac{\beta}{2} S)$. 
One can construct a one-parameter
set of actions which have no ${\cal O}(a^2)$ and ${\cal O}(a^4)$ corrections \cite{MP}:
\bea
\lefteqn{c_1 = (19- 55\  c_5)/9,\ \  c_2 =  (1- 64\  c_5)/9,}
\nonumber \\ 
\lefteqn{c_3 = (-64+ 640\  c_5)/45,\ \  c_4 = 1/5 - 2\  c_5 }
\label{coeffs}\eea
Here we take $c_5=1/20$ and $\beta=2.4$. 
We estimate the
contribution of all loops except the plaquette stochastically.
For each loop $l$ of sort $2\le i\le 5$ we introduce 
the auxiliary variable $\sigma_i(l)=0,1$; and rewrite the contribution of this 
loop to the Gibbs factor in the form 
\be
e^{-\frac{\beta}{2}S_{i,l}}\! =\!\!\!\!\!
\sum_{\sigma_i(l)=0,1}\!\! [\delta_{\sigma_i(l),0}
+\delta_{\sigma_i(l),1}(e^{-\frac{\beta}{2}S_{i,l}}-1)]
\ee
After each $N_i$ updates of the fields $U$ we update the $\sigma$ fields of sort $i$. 
We measure the average values of $\sigma_i$, as listed in Table \ref{t1}.
They are quite small, so one avoids the computation
of almost all of the extended "staples" in the $U$ update. 

\vspace{-.4cm}
\begin{table}[htb]
\caption{\small Average value of $\sigma_i$ field
for each loop of sort $i$.}
\label{t1}
\begin{tabular}{|c|c|c|c|c|} \hline
loop & 1x2 & 1x3 & 2x2 & 3x3 \\
\hline
$\langle \sigma \rangle $& 0.0753 & 0.0199 & 0.0202 & 0.0018\\
\hline
\end{tabular}
\end{table}
\vspace{-.4cm}

The efficiency of NMC is estimated by comparing it with the non-noisy
updating procedures (heatbath, overrelaxation) which are commonly used for the simulation of  
actions like (\ref{5Li}).We label these usually 
applied techniques with the collective name "Usual Monte Carlo" (UMC), 
to contrast with NMC.

For NMC the average computational cost per one update of the $U$ fields 
on the entire lattice in units of matrix (link) multiplications is 
\bea
\lefteqn{t_{tot}^{NMC}= t_U^{NMC}+\sum_{i=2}^{5}\frac{t_{\sigma_i}}{N_i}=}
\nonumber\\
=\lefteqn{t_U^{pl}+6V\sum_{i=2}^{5} P_i s_i (\frac{1}{N_i}+P_i  \langle \sigma_i\rangle )}
\label{nmctime}\eea 
where $t_U^{pl}$ is the update cost for the elementary plaquette action,  
$P_i$ is the perimeter of loop $i$, $s_i$ is
a symmetry factor: $s_i=1$ for square loops and $s_i=2$ for rectangular 
loops (see Ref.\cite{BF} for details).
The computational cost for 
UMC per $U$ update is approximately equal to the 
r.h.s. of expression  (\ref{nmctime}) in the limit $N_i\rightarrow\infty$ and
$\langle \sigma_i\rangle \rightarrow 1$:
\be
t_{tot}^{UMC}=t_U^{pl}+6V\sum_{i=2}^{5} P_i^2 s_i   
\label{umctime}\ee

\begin{table*}[htb]
\caption{Integrated autocorrelation times for average loop traces in units of $U$
updates for the UMC algorithm (first column), and the NMC algorithm with different frequencies of
$\sigma$ updates for 1x2, 1x3, 2x2 and 3x3 loops (other columns). The last row presents the 
naive gain for the NMC algorithm (\ref{gain}).
The real gain is given by eq.(\ref{real}) and depends on the observable.
From the values of $\tau_{int}$ below, the real gain is ${\cal O}(4-6)$.}
\label{t2}
\begin{tabular}{|c|c|c|c|c|c|c|c|c|c|} \hline
number of $U$  & UMC    &1  &5  &5  for 1x2 & 10&10 for 1x2    &20 &30 &40 \\
updates per    & no     &for&for&15: 1x3,2x2&for&30: 1x3,2x2&for&for&for\\
1 $\sigma$ update  &$\sigma$&all&all&105 for 3x3&all&210 for 3x3   &all&all&all\\
\hline
$\tint$ (1x1)&0.7(1)&1.9(2)&2.3(1)&2.5(2)&3.1(2)&3.2(2)&4.3(4)&4.5(4)&3.8(2)\\
\hline
$\tint$ (1x2)&0.8(1)&2.6(3)&2.8(2)&3.2(2)&4.3(4)&3.9(3)&5.2(4)&5.6(5)&5.7(4)\\
\hline
$\tint$ (1x3)&0.8(1)&2.7(3)&2.8(2)&3.2(2)&4.3(4)&3.9(3)&5.1(4)&5.4(5)&5.4(4)\\
\hline
$\tint$ (2x2)&1.0(1)&3.4(5)&3.3(3)&3.7(3)&4.7(4)&4.7(3)&5.3(4)&5.8(6)&5.7(4)\\
\hline
$\tint$ (2x3)&1.4(3)&4.2(7)&3.8(3)&4.2(3)&5.5(5)&5.7(4)&5.7(5)&6.2(6)&6.3(5)\\
\hline
$\tint$ (3x3)&1.8(4)&5.0(8)&4.5(5)&5.0(5)&5.8(6)&6.4(5)&5.9(5)&6.3(6)&6.3(5)\\
\hline
$r^{naive}_{gain}$&1&7.2&14.8 &18.3&17.9 &20.6&18.3&20.8&21.1\\
\hline
\end{tabular}
\end{table*}

The naive gain in efficiency from using NMC is given by the ratio between the 
costs (\ref{umctime}) and (\ref{nmctime}):
\be 
r^{naive}_{gain}=
\frac{t_U^{pl}+6V\sum_{i=2}^{5} P_i^2 s_i}
{t_U^{pl}+6V\sum_{i=2}^{5} P_i s_i (\frac{1}{N_i}+P_i  \langle \sigma_i\rangle )}
\label{gain}\ee
One should also take into account the increase of autocorrelation
times coming from the introduction of variables $\sigma$ in the NMC algorithm, 
so the real gain is
\be
r^{real}_{gain}\equiv r^{naive}_{gain} *  \frac{\tint^{UMC}}{\tint^{NMC}}\quad .
\label{real}
\ee
where the second factor depends on the observable under consideration.

It is not practical to keep the same updating frequencies
$1/N_i$ for all sorts $i$ of loops. In order for the work in the $\sigma$ and in the
$U$ updates coming from loops of sort $i$ to remain comparable, one should keep the updating
frequencies $1/N_i$ proportional to $\langle \sigma_i\rangle $:
\be
\frac{1}{N_i} \;\sim  \; P_i \langle \sigma_i\rangle          
\label{var}\ee
Due to the small influence of weakly
coupled terms on the dynamics of the system, one can expect only insignificant changes in
the autocorrelation behavior as $N_i$ increases. 

Table \ref{t2} gives an impressive demonstration of the benefits which come 
from using the NMC algorithm. 
For the runs where the updating frequencies for $\sigma$ fields are adjusted as per 
eq.(\ref{var}) we infer that the 'real gain' in computer time
is of order ${\cal O}(4-6)$. 
One can expect a much greater gain for more complicated
highly-improved actions.  

Let us now speculate on possibilities to use our algorithm to simulate a
Hamiltonian with a very large number of terms. A specific example we have in
mind is the case of full QCD, where the measure is, for 2 flavors of Wilson
quarks:
\be
\frac{1}{Z} e^{-S_g(U)} ~~ det^2({\bf 1} - \kappa M(U))
\label{measure}
\ee
where $S_g$ is the local gauge action, $M(U)$ is a hopping matrix connecting
nearest neighbours on a $4d$ hypercubic grid, and $Z$ normalizes the distribution.
The determinant can be turned into $exp(Tr(Log({\bf 1} - \kappa M(U))))$, 
then the logarithm expanded around 1,
giving the loop expansion of the measure above:
\be
\frac{1}{Z} e^{-S_g(U) - 2 \sum_{l=4}^\infty \frac{\kappa^l}{l} Tr M(U)^l} \quad .
\label{loops}
\ee
$Tr M(U)^l$ can be represented as a sum over all closed  non-backtracking loops of 
length $l$ on the $4d$ hypercubic lattice. The number of types of contributing loops 
$n_l$ is bounded by $7^l$.
Although this upper bound is not saturated, it is clear that the multiplicity
of terms of a given length $l$ grows exponentially:
\be 
 n_l \sim F_1(l) \;\alpha^l;\qquad \alpha<7
\ee
where $F_1(l)$ is a rational function of $l$, and $\alpha^l$ is the leading 
exponential behavior of the number of loops of length $l$ in the limit of large $l$.
At first sight, it seems
that sampling numerically the distribution (\ref{loops}) is a disastrous idea:
the action contains an infinite number of terms, of exponentially growing 
multiplicity. Instead, other strategies are being used, based on the 
transformation of the determinant (\ref{measure}) into a Gaussian integral.

Nevertheless, the coupling $\frac{\kappa^l}{l}$ decreases exponentially as
$l$ increases. Therefore, the auxiliary variables $\sigma_l$ associated in
our approach with various loops of length $l$ will take value 0 almost always.
One gets:
\be 
\langle \sigma_l\rangle \; \sim c_l \sim F_2(l) \; k^l \gamma^l
\ee
where the exponentially growing factor $\gamma^l$ comes from the average 
trace of Dirac matrices along the loops of length $l$. 

If one arranges the updating frequencies for each loop $i$ as per 
eq.(\ref{var}), one can expect that the average computer time needed
for estimating the contribution of all loops of length $l$ behaves as
\be
t_l \sim \; n_l l^2  \langle \sigma_l\rangle  \; \sim  F(l) * (\alpha\gamma\kappa)^l
\ee
For $\;\kappa < \kappa_{ca}=\frac{1}{\alpha\gamma}\;$
the computational cost $t_l$ decreases exponentially with $l$
and the total computational cost of the algorithm $t=\sum_l t_l$ 
converges to some finite value. 
Our rough estimation from fitting $\; n_l\;$ and average trace of Dirac matrices
in the interval $4\le l\le 12$ gives $\;\alpha\approx 5.4;$ $\gamma\approx 1.4$,
and therefore $\kappa_{ca} \approx 0.13$. 

In the regime $\kappa <\kappa_{ca}$ we are in an interesting situation 
where the influence of very large loops
is negligible because their associated coupling in the effective action is
extremely small. Therefore, truncating the loop expansion above a certain
order will introduce a statistically unobservable bias. Equivalently, one
can freeze the associated $\sigma$ variables at the value zero, or update
them with arbitrarily low frequency. In spite of this extremely (or infinitely)
slow dynamic mode of the $\sigma$'s, the dynamics of the gauge fields are not
affected.
Note that the cost of our algorithm grows linearly with the volume $V$ of the
system. This is better than alternative approaches to the simulation of
full QCD: Hybrid Monte Carlo (cost $\propto V^{5/4}$) and MultiBoson
(cost $\propto V(Log V)^2$) \cite{Bielefeld}. In addition the stepsize, or 
typical change at each update of a gauge link $U$, does not seem restricted
a priori for small quark mass, unlike in the two alternative approaches above.

A less speculative use of our algorithm for full QCD consists of truncating
the loop expansion eq.(\ref{loops}) to some order $l_{max}$, and representing
the higher orders with the MultiBoson approach \cite{MB}. This strategy,
called ``UV-filtered MultiBoson'', has already been used successfully
\cite{Forcrand}. However, in Ref.\cite{Forcrand} the loop expansion is truncated 
to its lowest term $l=4$, because the exact evaluation of larger loops is too
time-consuming. With our stochastic approach, these larger loops can be 
estimated at low cost. We expect this composite strategy to be particularly
efficient.

{\bf Acknowledgments:}
T.B. was supported by
INTAS 96-0370 and RBRF 99-01-00190.

\end{document}